\documentclass[10pt]{article}
\usepackage[left=1.2in,top=1.2in,bottom = 1.2in,right=1.2in]{geometry}
\usepackage{amsmath}
\usepackage{amssymb}
\usepackage{setspace}
\usepackage{mathrsfs}
\usepackage[english]{babel}
\usepackage{fontenc}
\usepackage{graphicx}
\usepackage{subfig}
\usepackage{psfrag}

\usepackage{color}

\usepackage{graphicx}

\newtheorem{theorem}{Theorem}[section]

\newtheorem{remark}[theorem]{Remark}

\newcommand{\req}[1]{(\ref{#1})}
\newcommand{\bm}[1]{\mbox{\boldmath{$#1$}}}

\def\zl{{\zeta}}
\def\ztu{{{\eta}_1}}
\def\ztl{{{\eta}_2}}

\def\ouu{{{\overline u_1}}}
\def\ouuj{{{\overline u_j}}}
\def\oul{{{\overline u_2}}}

\def\uu{{u_1}}
\def\uuj{{u_j}}
\def\ul{{u_2}}

\def\hu{{h_{1}}}
\def\hl{{h_{2}}}

\begin{document}

\date{}
\title{An inertia `paradox' for  incompressible stratified  Euler fluids}
\author{R. Camassa$^1$ S. Chen$^1$ G. Falqui$^2$ G. Ortenzi$^2$ M. Pedroni$^3$}

\maketitle

\begin{center}
$^1$Carolina Center for Interdisciplinary Applied Mathematics, Department of Mathematics, University of North Carolina, Chapel Hill, NC 27599, USA 
\\
$^2$Dipartmento di Matematica e Applicazioni, Universit\`a di Milano-Bicocca,  Milano, Italy
\\
$^3$Dipartimento di Ingegneria dell'Informazione e Metodi
Matematici, Universit\`a di Bergamo,
Dalmine (BG),  Italy
\end{center} 

\doublespacing

\begin{abstract}
The interplay between incompressibility and stratification can lead to non-conservation
of horizontal momentum in the dynamics of a stably stratified incompressible Euler fluid filling 
an infinite horizontal channel between rigid upper and lower plates. Lack of conservation 
occurs even though in this configuration only vertical external forces act on the system. This apparent paradox
was seemingly first noticed by Benjamin (\textit{J. Fluid Mech.}, vol. 165, 1986, pp. 445-474) in his classification of the invariants by symmetry groups with the Hamiltonian structure of the Euler equations in two dimensional settings, but it appears to have been largely ignored since. By working directly with the motion equations, the paradox is shown here
to be a consequence of the rigid lid constraint coupling through incompressibility with the infinite inertia of the far ends of the channel, assumed to be at rest in hydrostatic 
equilibrium. Accordingly, when inertia is removed by eliminating the stratification, or, remarkably,  by  using the Boussinesq approximation of uniform density for the inertia terms, horizontal
momentum conservation is recovered. This interplay between constraints, 
action at a distance by incompressibility, and inertia is illustrated by layer-averaged exact results, two-layer long-wave models, 
and direct numerical simulations of the incompressible  Euler equations with smooth stratification.
\end{abstract}

\section{Introduction}
\label{intro}

Among the many areas of classical mechanics, fluid dynamics arguably holds a special 
distinction for being a rich source of the sort of paradoxes that often arise from simplifying limit assumptions. 
Thus, for instance, the limit of zero viscosity gives rise to D'Alembert's paradox on 
the drag experienced by rigid bodies moving in ideal fluids, 
while the opposite 
limit of dominating viscous stresses leads to the  Stokes or Whitehead paradoxes of 
unphysical divergences for the same problem. 

This work focuses on an effect that could also be viewed as paradoxical: horizontal momentum 
conservation is violated in the dynamics of a stratified ideal fluid filling an infinite horizontal channel between
rigid bottom and lid boundaries, starting from localized initial conditions, even though the only external forces acting on the system are vertical 
(gravity and constraint forces from the horizontal boundary)
and the fluid is free to move laterally. Of course, even for an inviscid fluid, lateral boundaries could lead to horizontal forces by action-reaction mechanisms due 
to the constrained motion, and so horizontal momentum conservation cannot in general be expected
to hold for a stratified Euler fluid filling a finite domain enclosed by a rigid boundary. However, we shall see below that 
for a domain extending horizontally to infinity the infinite inertia possessed by the far fluid at rest  acts as an effective lateral boundary, giving rise to violation of horizontal momentum 
conservation. While stratification is necessary for  creating the relative inertia of the lateral 
fluid at rest, a subtlety of this effect is that  incompressibility is also required to transmit 
forces arising from finite-range motion 
instantaneously all the way to infinity. Accordingly, the ``light-cone" provided 
by the maximum speed of propagation of internal baroclinic modes gives a rough estimate of the boundary of
the exterior region that can be considered as contributing to an effective-wall lateral confinement. 

To the  best of our knowledge, this limiting behaviour in the dynamics of a stratified fluid has not been given 
much attention in the literature. Benjamin (1986)
appears to be the first to point out this curious property,
in the course of his investigation on symmetries and Hamiltonian structures of the stratified, incompressible two--dimensional Euler equations. 
In particular, 
Benjamin shows that the invariant  generally  associated 
with translational symmetry is the fluid's impulse rather than  its momentum.

This Hamiltonian approach is compact and elegant, and its applications 
certainly deserve further study. Nonetheless, 
the physical mechanisms responsible for the dynamics seem
to be more transparent by a direct approach with the simplest configuration of 
a two-layer fluid.  This configuration has the added advantage of leading naturally into 
reliable models when long-wave asymptotics applies. A further advantage of the 
direct approach is that it can be immediately extended to three-dimensional settings for fluid domains 
between horizontal rigid planes. 
Admittedly, the effect considered here can be viewed as small, because momentum conservation is  recovered as the size $\Delta \rho$ of the density range  (which in practical cases such as 
water stratified with heat or salt, is typically $\Delta \rho /\rho \simeq 10^{-2}$)vanishes.  Of course,  the  effect also relies on the abstract setup of infinite 
rigid bounding surfaces.  Nonetheless, we think that 
this limiting case is of conceptual importance for a proper
understanding of the dynamics of 
the incompressible limit for density-stratified fluids. 

The paper is organized as follows. In \S2 we first derive balance laws that imply the paradox for incompressible stratified Euler equations in an infinite channel, without approximations. Next, we show that the paradox
remains in a two-layer fluid in the hydrostatic (dispersionless) non-Boussinesq approximation. In this simpler setting  an explicit formula for the interface pressure can be derived. In \S3, we show 
how the paradox can arise via direct numerical simulations of stratified incompressible Euler equations.

\section{Layer averaged Euler equations}
While the inertia effects that we focus upon here arise with general smooth 
stratifications, we work first with two-layer fluids. This setup is the 
most convenient for developing long-wave  models, which  can further 
illustrate the inertia effect by allowing explicit formulae to be derived.  
Similarly, the restriction to a single horizontal dimension 
is not essential, and our conclusions (and formalism) work 
for the full three-dimensional case of a horizontal fluid between infinite top and bottom 
rigid bounding plates. 
We choose to work with layer-averaged equations, which of course
can be formulated independently of the assumption of 
stacked homogeneous layer stratification. 

The dynamics of an inviscid and incompressible fluid 
stratified in layers of uniform density $\rho_j$
is governed by the Euler equations for 
the velocity components $(u_j,w_j)$ 
and the pressure $p_j$,  in two dimensional Cartesian coordinates $(x,z)$, 
\begin{eqnarray}
       {u_j}_x+{w_j}_z&=&0,
\label{conti}
\\
       {u_j}_t+u_j {u_j}_x+w_j {u_j}_z&=&
       -{p_j}_x/\rho_j,\label{heuler}
\\
       {w_j}_t+u_j {w_j}_x
       +w_j{w_j}_z&=&-{p_j}_z/\rho_j-g,
\label{veuler}
\end{eqnarray}
where $g$ is the gravitational acceleration and
subscripts with respect to space and time represent
partial differentiation.
In a two-fluid system,  $j=1$ ($j=2$) stands
for the upper (lower) fluid,
and $\rho_1\leq \rho_2$  must be assumed 
for stable stratification.

\begin{figure}
\centering
{\includegraphics[width=12cm,height=5cm]{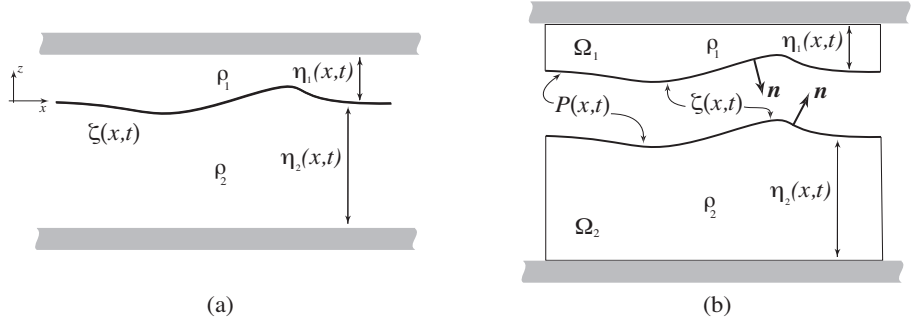}}
\caption{ (a) Two-layer fluid setup and relevant 
notation. (b) The domains for computation of momentum balance}\label{pressure_setup}
\end{figure}

For a channel  with upper and lower rigid surfaces (see figure~\ref{pressure_setup}a for 
the setup and relevant notation)
the kinematic boundary conditions are 
\begin{equation}
       w_1(x,\hu,t)=0\,,\qquad w_2(x,-\hl,t)=0\, ,\label{ubc}
\end{equation}
where $\hu$ ($\hl$) is the undisturbed thickness of the upper (lower)
fluid layer, respectively.
The boundary conditions at the interface $ z=\zl(x,t)$ are the continuity of normal velocity and pressure
\begin{equation}
       \zl_t+u_1\zl_x=w_1,\quad 
       \zl_t+u_2\zl_x=w_2,\quad
        p_1=p_2\equiv P
       \qquad {\rm at}\quad z=\zl(x,t),
       \label{lbc}
\end{equation}
where $\zl(x,t)$ is the displacement of the interface from the equilibrium configuration surface $z=0$ 
and $P(x,t)$ denotes the interfacial pressure.
As to the lateral boundary conditions, a set of particular interest physically is the one that corresponds 
to localized initial data, i.e., the fluid
is quiescent at infinity. This would require 
\begin{equation}
\zeta(x,\cdot) \to 0 \, ,  \quad {\bm u}_j(x,\cdot,\cdot) \to 0\, ,   \quad j=1,2 \, , 
\qquad \hbox{as} \quad |x| \to \infty \, ,
\label{bcuz}
\end{equation}
sufficiently fast, which in turn implies that at infinity hydrostatic equilibrium applies, 
\begin{equation}
{p_j}_z+\rho_j g=0, \quad j=1,2\, , \qquad \Rightarrow \qquad  p_j=-g \rho_j z+P\, , 
\qquad \hbox{as} \quad |x| \to \infty \,  . 
\label{bcup}
\end{equation}

In what follows we rewrite the Euler system~(\ref{veuler}) in terms of 
layer averages (see e.g. Wu 1981). 
(For a smoothly stratified fluid, this is equivalent to singling out an intermediate 
level set of constant density $z=\zeta(x,t)$ and carrying similar 
manipulations since such a set will always be a material surface.)
We define the layer-mean quantities $\bar f$ as
\begin{equation} 
       \bar f(x,t)
       \equiv{1 \over \eta_j} \int_{[\eta_j]}
       f(x,z,t) {\rm d}z \,,  
       \label{defint}
\end{equation}
where $\eta_j$ are the layer thicknesses $\eta_i\equiv h_j+(-1)^j \zeta$, and, abusing notation a little by not differentiating overbars with respect to lower or 
upper layer, the intervals of integration $[\eta_j]$ are $z\in(\zeta,h_1)$ for the upper  and 
$z\in (-h_2 ,\zeta)$ for the lower layers, respectively. 
Vertically integrating  (\ref{conti})--(\ref{heuler}) across the layers and 
imposing  the boundary conditions  \req{ubc}--\req{lbc}
yields the layer-mean equations for the upper (lower) fluid 
\begin{eqnarray}
       {\eta_j}_t+\big (\eta_j\ouuj\big )_x 
       &&=0,
\label{meancu}
\\ 
      \rho_j (\eta_j\ouuj)_t
       +\rho_j\big (\eta_j\overline {\uuj\uuj}\big )_x 
       &&=
       -(\eta_j\overline {{p_j}})_x +(- 1)^j \zl_x P\,, \qquad j=1,2\, . 
\label{meanhupper}
\end{eqnarray}
(We use the notation $\overline {\uuj\uuj}$ here and in similar formulae below instead of 
the equivalent $\overline {\uuj^2}$ because the latter applies only to the 
two dimensional case, whereas the  former can be used for three dimensions as well, 
upon interpreting the horizontal velocity product as a two-tensor and replacing the $x$-derivative by 
a divergence over the horizontal variables.)

For incompressible, inviscid fluids under a body-force density $\bm{f}(\bm{x},t)$  in a domain $\Omega$, the momentum 
balance in Eulerian form is expressed by 
\begin{equation}
{d \bm{\Pi}\over dt}\equiv {d\over dt} \int_\Omega \rho \bm{u} \, dV=
\int_\Omega {\partial \over \partial t}(\rho \bm{u}) \, dV+
\int_\Omega \hbox{div} (\rho \bm{u}  \bm{u} )\, dV =-\int_{\partial \Omega} p \bm{n} \, dA \, + \, 
\int_\Omega \rho \bm{f} \, dV \, , 
\label{postulate}
\end{equation}
where $\bm{n}$ is the outward normal to the surface $\partial \Omega$, and $dV$, $dA$ denote the
volume and area elements, respectively. 
Layer averages are just a local version of the integral form of the horizontal momentum balance
for each layer (see figure~\ref{pressure_setup}b), which can be expressed  by integrating 
equations~\req{meanhupper} 
over some $x$-interval $L_-\leq x \leq L_+$. We have 
\begin{equation}
{d {\Pi_1}_j \over dt}\equiv {d \over dt} \int_{L_-}^{L_+} \hspace{-10pt} \rho_j \eta_j \ouuj \, dx +
\left. \rho_j \eta_j\overline {\uuj\uuj} \right|_{L_-}^{L_+}=
-\left. \eta_j\overline {{p_j}} \right|_{L_-}^{L_+}+(-1)^j\int_{L_-}^{L_+} \hspace{-10pt} \zeta_x P \, dx \, ,
\label{postulatej}
\end{equation}
for the upper ($j=1$) and lower ($j=2$) layer respectively, since the 
outward normals along the interface are $\bm{n}\propto (\pm \zeta_x,1)$, and neither the pressure  at the rigid  horizontal surfaces or the external gravity field contribute horizontal components of 
forces. 

In hydrostatic equilibrium, the layer-mean pressures are 
\begin{equation}
\overline {{p_j}}=(-1)^j g \rho_j {\eta_j \over 2} +P \, , \quad j=1,2\, .
\label{hydropr}
\end{equation}
Hence, by a suitable definition of the limit procedure $L_\pm \to \pm \infty$,  the lateral  equilibrium boundary conditions imply  that for each infinite upper and lower layer the horizontal momenta are conserved if and only if
\begin{equation}
-h_1 \left. P\right|_{-\infty}^{+\infty}-\int_{-\infty}^{+\infty} \hspace{-2pt} \zeta_x P \, dx =0 \, , 
\qquad 
- h_2 \left. P\right|_{-\infty}^{+\infty}+\int_{-\infty}^{+\infty} \hspace{-2pt} \zeta_x P \, dx = 0 \, , 
\label{momhozcon}
\end{equation}
at all times, that is, if 
\begin{equation}
\int_{-\infty}^{+\infty} \hspace{-2pt} \zeta_x P \, dx =0 \qquad \hbox{and} \qquad 
\left. P\right|_{-\infty}^{+\infty}=0 \, . 
\label{momhozconred}
\end{equation}
(These relations are precisely the ones encountered in the study of 
single-layer fluids when an external pressure distribution is applied to their 
free surface.)

Summing up the two momentum equations \req{meanhupper} 
(for $j=1,2$)
yields the mean 
layer balance law for the total momentum of the fluid 
\begin{equation}
\partial_t\Big( \rho_1(\ztu\ouu)+ \rho_2(\ztl\oul)\Big)=-\partial_x 
      \Big(\rho_1\big (\ztu\overline {\uu\uu}\big )
       +\rho_2\big (\ztl\overline {\ul\ul}\big )+\ztu\overline {{p_1}}+\ztl\overline {{p_2}}
       \Big)\, .
       \label{momh}
\end{equation}
By action and reaction  the contribution from the pressure at the interface $P(x,t)$ drops from the 
balance~\req{momh} as well as from the integral version 
of the total horizontal momentum. Thus,  the condition for total momentum conservation
is that $\left. P\right|_{-\infty}^{+\infty}=0$, since~\req{hydropr} with~\req{postulatej} in this limit yields
$$
{d \Pi_1 \over d t}={d {\Pi_1}_1\over dt }+{d {\Pi_1}_2\over dt }=-(h_1+h_2)\left. P\right|_{-\infty}^{+\infty} \, . 
$$
At first sight, for localized displacements and velocities,  it might not be clear how the asymptotic values of the interfacial pressure could be 
different from plus to minus infinity,  as the hydrostatic equilibrium is identical at both ends and the interfacial 
pressure simply keeps track of the overall constant of integration up to which pressure is defined. 
For a free upper surface, this constant  is usually set by the atmospheric pressure;  if 
this is assumed to be uniform,  no pressure jump can occur. However, a system with a rigid lid is constrained, and reaction forces 
can develop in response to the constraint. 
Thus, we now focus on the consequences of the rigid lid constraint $\eta_1+\eta_2=h_1+h_2$. 
The continuity equations \req{meancu} 
imply 
\begin{equation}
\label{def-Q}
\big(\eta_1 \ouu +\eta_2 \oul \big)_x=0 \quad 
\Rightarrow \quad \eta_1 \ouu +\eta_2 \oul  \equiv Q(t) \, ,
\end{equation}
that is,  
the volume  flux $Q$ through the channel can only be a function of  time. 
Dividing the momentum equations \req{meanhupper} 
by the respective densities and summing the resulting equations  yields 
\begin{equation}
\partial_x \Big(\ztu\overline {\uu\uu}
       +\ztl\overline {\ul\ul}+{1\over \rho_1}\ztu\overline {{p_1}}+{1\over \rho_2}\ztl\overline {{p_2}}\Big)
       =\Big({1\over \rho_2}-{1\over \rho_1}\Big)\zeta_x P  -\dot{Q}\, . 
\label{pinterface}
\end{equation}
With the far-field zero boundary conditions on the velocities, which  implies $
Q(t)=0
$ 
at all times, 
equation~\req{pinterface}  can be interpreted as an expression that determines the (unknown) 
interfacial pressure $P(x,t)$ in terms of the divergence of the layer-mean quantities. 
By integrating in $x$ and taking into account the boundary conditions~\req{bcuz}-\req{bcup} we
obtain
\begin{equation}
\Big({h_2 \over \rho_2}+{h_1 \over \rho_1}\Big)\left. P\right|_{-\infty}^{+\infty}=
\Big({1\over \rho_2}-{1\over \rho_1}\Big)  \int_{-\infty}^{+\infty} \zeta_x P \, dx \, ,
\label{deltap_intp}
\end{equation}
which shows that unless the surface integral of the pressure along the interface at the right-hand-side of~(\ref{deltap_intp})
vanishes, 
or the layers have the same density, the extremal 
values of the interfacial pressure will in general be different. 
The equivalent expression 
$$
\rho_2\Big(h_1 \left. P\right|_{-\infty}^{+\infty}  + \int_{-\infty}^{+\infty} \zeta_x P \, dx \Big)
=
-\rho_1\Big(h_2 \left. P\right|_{-\infty}^{+\infty}  -\int_{-\infty}^{+\infty} \zeta_x P \, dx \Big) \, 
$$
shows that if one of the two conditions in~\req{momhozcon} is satisfied, i.e., horizontal momentum of one of the layers is conserved, the other will 
be as well, as the surface pressure integral is linked to the difference of asymptotic interfacial 
pressure by the rigid lid constraint. Thus,  conservation of the horizontal momentum of just one of 
the two layers implies conservation of the total horizontal momentum of the fluid. On the other hand, with 
nonzero  surface pressure 
integral along the interface total horizontal momentum will change with time, i.e., the bulk of the fluid will  in general undergo accelerations. 
Horizontal momentum is always conserved if the fluid is homogeneous, $\rho_1=\rho_2$, 
as~\req{deltap_intp} shows that in this case interfacial pressure forces cannot 
add up to provide a total pressure gradient between the far ends of the channel.  
Perhaps more notable is 
the effect of the Boussinesq approximation of taking $\rho_1=\rho_2$  in front of the inertial terms 
(cf. Boonkasame \& Milewski, 2011 for 
an analysis of the interplay between interfacial pressure and flux in the non-Boussinesq case 
and of the stability properties of the long-wave regime). 
Just as in the case of 
homogeneous-density fluid, build-up of pressure jump $\left.P\right|_{-\infty}^{+\infty}$ from interfacial pressure cannot occur
in this case:
taking the Boussinesq approximation in,  e.g., equation~\req{momh},  and applying the constraint $Q(t)=0$ sets 
the right-hand side of that equation to zero, so that $\dot{\Pi}_1=0$, which in turn implies $\left. P\right|_{-\infty}^{+\infty}=0$. 
Hence total as well as individual 
layer momenta are always conserved in the Boussinesq approximation for two-layer channel flows 
with far-field hydrostatic equilibrium boundary conditions. 

It remains to be seen if states of the fluid leading to a nonzero interfacial integral
at the right-hand of equation~(\ref{deltap_intp}) 
can develop during the evolution governed by the Euler equations (even for a general smoothly stratified fluid). 
A convenient starting point is offered by a choice of 
initial conditions corresponding to zero velocity and a local deformation of density level sets away from the (flat) ones for hydrostatic equilibrium. 
This is the choice of initial data used in the numerical simulations below, where in particular we take $x$-antisymmetric initial deformations. 
As we will see, during the subsequent evolution, this choice leads to an analogue for a finite domain of time variation of horizontal momentum for 
the infinite channel. 
The numerical simulations will be performed with near-two-layer configurations, and with initial data which are slowly varying in $x$. 
For such a case explicit expressions (not readily available in the general case) for the quantities in equation~(\ref{deltap_intp}) can be derived 
approximately using long-wave asymptotics. 
\subsection{Shallow water models}
At leading order in a long-wave asymptotic expansion (see e.g. Yih, 1980), the hydrostatic approximation for the pressures
holds  throughout the fluid domain, 
not just as far-field boundary conditions. This can be used to derive a 
closed form expression for the interfacial pressure in equation~(\ref{pinterface}). 
The result is expressed in terms of well known two-layer (five-equation, dispersionless) shallow-water 
model (see e.g.~\cite{tabak-2004}).
We have 
$$
\partial_x \Big (\ztu\overline {\uu\uu}
       +\ztl\overline {\ul\ul}+g \zeta (h_1+h_2) +g{h^2_2-h_1^2 \over 2}+\Big({\eta_1\over \rho_1}+{\eta_2\over \rho_2}\Big)P\Big)
       =\zeta_x  \Big({1\over \rho_2}-{1\over \rho_1}\Big) P \, , 
$$
so that, with the identities $\zeta_x=-{\eta_1}_x={\eta_2}_x$ used in the right-hand side of this expression, 
$$
\partial_x \big (\ztu\overline {\uu\uu}
       +\ztl\overline {\ul\ul}+g \zeta (h_1+h_2) \big)
       = -P_x \Big({\eta_1\over \rho_1}+{\eta_2\over \rho_2}\Big) \, .
$$
Upon splitting the average of products into the products of averages, this coincides with 
the expression derived from the five-equation model, and yields 
\begin{equation}
\left.P\right|_{-\infty}^{+\infty} 
       =-
       {\rho_1\rho_2}\int_{-\infty}^\infty {(\ztu\overline {\uu}^2
       +\ztl\overline {\ul}^2 )_x \over \rho_1\eta_2+\rho_2\eta_1}dx \, .
\label{deltaPleadingorder}
\end{equation}
Here a term with the factor $g(h_1+h_2) \zeta_x$ has been dropped because the denominator is only a (linear) function of $\zeta$ 
thus making the ratio a perfect $x$-derivative, vanishing when the boundary conditions on $\zeta$ are applied. 
Thus,  at leading order
total horizontal momentum conservation requires the extra constraint on the choice 
of initial data that make the above integral vanish, which is manifestly not satisfied for general
functions $\overline{u}_j$ and $\eta_j$. Note that if $\rho_1=\rho_2$ the denominator in the 
integrand in~\req{deltaPleadingorder} becomes a constant, making the integral null on account of 
the velocity boundary conditions. 

Finally, we remark that equation (\ref{deltaPleadingorder}) shows that the symmetries of the system with respect to the horizontal variable allow the identification of a large class of solutions compatible with momentum conservation (in the hydrostatic approximation). Indeed, it is easy to check that if initially $\eta_1$, $\eta_2$ are even functions and $\overline{u}_1$, $\overline{u}_2$ are odd functions  with respect to $x$, then these symmetries are preserved by the evolution of the system. For such solutions, (\ref{deltaPleadingorder}) shows that the (null) horizontal momentum is conserved. However, generic initial 
conditions not in this class can be shown to evolve to non-zero $\left.P\right|_{-\infty}^{+\infty}$, even starting from null values of this pressure jump, or, 
remarkably, even when the velocities are chosen to be initially zero. For this latter case, this can be seen by looking at the higher order 
dispersive (non-hydrostatic) corrections to the shallow-water model  as reported in Choi \& Camassa (1999). 
At $t=0$ with zero initial velocities these corrections modify equation~(\ref{deltaPleadingorder}) as
\begin{equation}
 \left.P\right|_{-\infty}^{+\infty}=\frac{1}{3}\int_{-\infty}^{+\infty} \frac{ \left( \eta_1^3 {\ouu}_{xt}+\eta_2^3 {\oul}_{xt} \right)_x }{\eta_1/\rho_1 +\eta_2/\rho_2}\ dx \, , 
\label{dispers_deltap}
\end{equation}
which, by bringing into the integrand the time-derivatives of the velocities shows that the pressure jump can be 
non-zero even if the velocities are initially zero. In particular, antisymmetric initial displacements 
of the interface can lead to non-zero $P{|}_{-\infty}^{+\infty}$, whereas this pressure jump always vanishes
 for symmetric initial data. 

\begin{figure}
\centering
{\includegraphics[width=10cm]{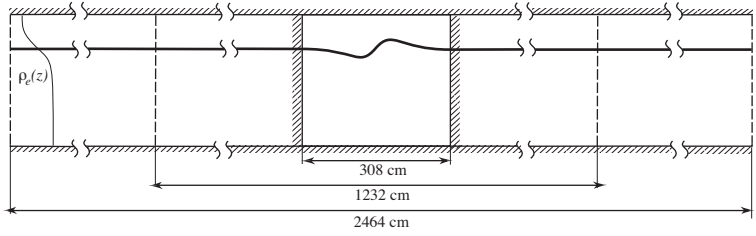}}
\caption{Sketch of the fluid test domain and its symmetrical padding  by wings of increasing length, doubling and quadrupling  the period as shown.}
\label{sketch_pad}
\end{figure}
\section{Numerical simulations}
The above discussion was conducted with laterally unbounded domains in mind. 
Of course, such an idealization cannot be used either in reality or in numerical studies. 
However, in this section 
we provide numerical evidence that the effective-wall lateral confinement, 
and hence  non-conservation of horizontal momentum, can occur in  {\it finite} 
domains,  due to the relative inertia
of a stratified, incompressible Euler fluid. First, we remark that, for domains bounded by rigid lateral walls, the finite-domain 
version of equation~(\ref{deltap_intp}) (obtained by writing $\pm L/2$ in place of $\pm \infty$) 
continues to hold; in the limit of the walls moving  to infinity we simply recover the hydrostatic 
balance as expressed by~(\ref{deltap_intp}). 
Next, consider the case of periodic boundary conditions in the periodic box $[-L/2,L/2]$. 
This requires $\left.P\right|_{-L/2}^{+L/2}=0$ and hence the horizontal momentum for 
the whole periodic domain is conserved. 
We focus on a subset of the fluid domain, henceforth referred to as the 
``test section," obtained by taking a (much) smaller interval embedded in the 
period (cf. figure~\ref{sketch_pad}). Within this test section, we apply localized initial conditions
 for velocity and pycnocline displacement, e.g., by requiring that the data
 have compact support on a small subset of the test section's region.  The analogue 
 of equation ~(\ref{deltap_intp}) for a periodic 
domain becomes an equation for the flux~$Q$, 
\begin{equation}
 L\dot{Q} = \left(\frac{1}{\rho_2}-\frac{1}{\rho_1}\right) \int_{-L/2}^{L/2} \zeta_x P\ dx.
\label{qt}
\end{equation}
Consider the limit $L\to \infty$ of this equation.  For definiteness, let $\zeta$ 
be a function with compact support and suppose that all the velocities are zero at $t=0$. 
The integral on the right-hand side will be bounded as $L\to\infty$ (assuming that 
$P$ remains bounded on finite domains), so that  
$\dot{Q} \sim L^{-1}$. 
Suppose the test section extends from $-A/2$ to $A/2$ and $\mathrm{supp}(\zeta)\subset [-A/2,A/2]$.
At $t=0$, after integrating (\ref{pinterface}) in the test section and eliminating $\dot Q$, we obtain 
\begin{equation}
 \left(\frac{1}{\rho_2}-\frac{1}{\rho_1}\right) \left(1-\frac{A}{L}\right) \int_{\mathrm{supp}(\zeta)} \zeta_x P\ dx =
\left(\frac{h_2}{\rho_2}+\frac{h_1}{\rho_1}\right) P \Big{|}_{-A/2}^{+A/2} \, .
\end{equation}
If we extend the test section to infinity with the double scaling limit $A,L \to \infty$ and $A/L \to 0$, the previous formula becomes (\ref{deltap_intp}). 
Though valid only at time $t=0$, this argument shows how the limit of infinite period for localized initial data can agree with  the pressure 
differential of the infinite channel in hydrostatic balance at infinity. 

We now explore numerically the time evolution of localized initial data under both periodic and rigid (impermeable) wall boundary conditions. 
In particular, we first compute  the evolution of the flux $Q(t)$ and horizontal momentum $\Pi_1(t)$ 
for the test section alone. We then compare the resulting time series with those from simulations from the same initial conditions in 
progressively longer channels {\it under periodic boundary conditions}, see figure~\ref{sketch_pad}. Thus, while the 
total horizontal momentum for these longer periodic channels is conserved, that computed only in the
embedded test-section will in general exhibit time dependence. Owing to the added 
inertia of the ``padding" wings bracketing the test section in the longer channels, we expect this time dependence 
to show some similarity with that of the walled-in test section. That is,  the 
added inertia acts as virtual walls, which could then approximate actual walls in the 
limit of an infinite periodic channel.

\begin{figure}
\centering
{\includegraphics[width=13cm]{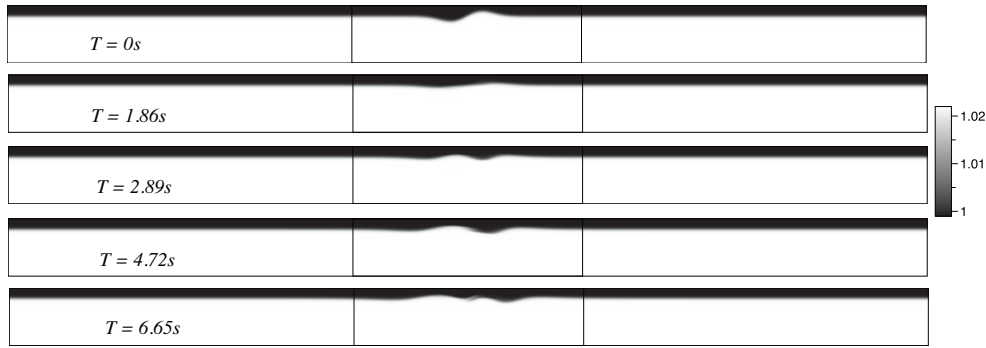}}
\caption{Density field from the numerical simulation of the evolution from  the initial data in the 1232 cm long tank with the center 308 cm test  section marked by the vertical lines.}
\label{dns}
\end{figure}
\begin{figure}
  \centering
  \includegraphics[width=13.5cm]{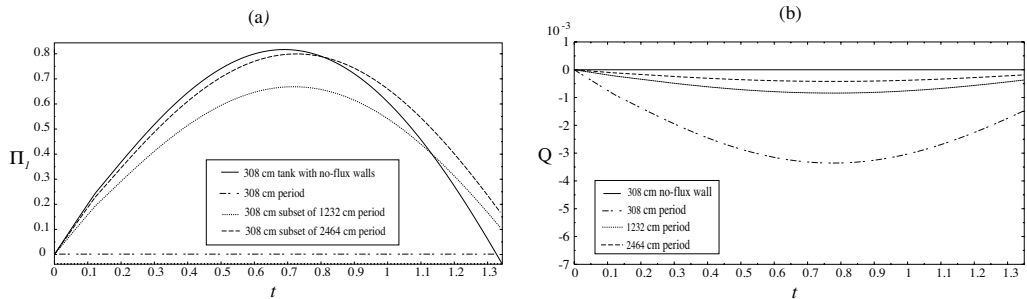}
\caption{(a) Horizontal momentum time evolutions for the test section embedded in progressively larger periodic domains, 
starting from the same initial condition. The solid line correspond to the rigid wall boundaries. (b) Time series of fluxes $Q(t)$ with respect to increasing period $L$, for the same cases as (a). The flux decreases as $1/L$ in response to the larger inertia of the channel ``padding" wings. 
}
\label{3tre}
\end{figure}

The details of our numerical simulations are as follows. 
The initial conditions in all our simulations (all performed using dimensional quantities, and translating the 
origin of the coordinates to the bottom) 
are chosen to be the antisymmetric interface displacement through
$\zeta_0(x)= h_2 +{x/ 2} \exp(-{x^2 / \sigma^2})$
together with zero initial velocities. This function displaces the smooth equilibrium density function $\rho_e(z)$ to give the initial condition $\rho_0$ (with obvious meaning of notation)
\begin{equation}
\rho_0(x,z)=\rho_{1}+{\rho_{2}-\rho_{1} \over 2}\left(1+\tanh\left[{\gamma}(\zeta_0(x)-z)\right]\right)\, ,\qquad 
z\in[0\,,H]\,.
\label{eq:strat_tanh}
\end{equation}
Here, $\sigma=30$ cm, $\rho_1=0.999$ g/cm$^3$, $\rho_2=1.022$ g/cm$^3$, $H=77$ cm, 
$h_2=62$ cm, and the thickness of the pycnocline (defined as the distance 
between density isolines corresponding to $10$\% and $90$\% of the total 
density jump) is set by the parameter~$\gamma=0.5$ to
correspond to about $4.5$ cm (all of these parameters are suggested by those typical for experiments with 
salt-stratified water). Notice that this choice of parameters gives  effectively  an initial condition of compact support, with the initial  departure from hydrostatic equilibrium for $|\rho-\rho_e|/\rho_e$ of order 
$10^{-10}$ at the boundary of the test section $x=\pm 154$~cm; this departure remains below $10^{-7}$ in all our runs.
The simulations (see figure~\ref{dns}) are performed using the numerical code VARDEN which solves the stratified incompressible Euler equations 
(for details see \cite{Almgren et al. (1998)}.) We typically use a square grid with $512$ points along the vertical, although we have run cases with doubled and half this resolution to assess convergence.
Figure~\ref{3tre}a shows the time series of the horizontal momentum of the test section for the 
walled-in configuration, and compares it to that computed with periodic boundary conditions with quadrupled and octupled 
periodic extensions. As can be seen, there is indeed a tendency for the longer channel to 
yield a momentum evolution closer to that of the walled section, for the initial (short) time 
displayed. As expected, later time evolution shows larger discrepancies but still with 
similar overall behavior and magnitudes. This is in rough agreement with the estimate
from the fastest baroclinic wave speeds, which for this parameter choice are of order $16$ cm/s, 
and with the horizontal scale of the initial condition with respect to that of the test section. 
For reference, we remark that the code maintains the total horizontal  momentum for the periodic 
channels close to zero (the initial value) with an error of order $10^{-3}$. 
Figure~\ref{3tre}b presents the time series of the flux $Q(t)$ for the same runs. 
The flux is computed at different 
$x$-locations, yielding the same value to within a relative error of $10^{-10}$ (thus further
validating the convergence of the code). 
As can be seen by the different curves, 
the flux appears 
to scale as the inverse of the channel length $L$, in agreement with expression~(\ref{qt}) for its 
initial time derivative. 
This can be taken as further evidence of the inertia provided by the padding wings (growing as $L$) 
which acts to oppose the fluid flux (recall that in the limit of an unbounded domain $Q\equiv 0$
due to the equilibrium at infinity). 
The inverse scaling with $L$
can be given 
further analytic interpretation. 
In fact, the analogue of~\req{deltaPleadingorder} for the leading-order hydrostatic (and hence dispersionless) long-wave 
approximation is 
\begin{equation}
\dot{Q} \int_{-L/2}^{+L/2} {1 \over \eta_1/\rho_1+\eta_2/\rho_2} \, dx 
+\int_{-L/2}^{+L/2} {(\eta_1 \overline{u_1}^2 +\eta_2 \overline{u_2}^2)_x
\over \eta_1/\rho_1+\eta_2/\rho_2} \, dx =0\, . 
\label{qtsw}
\end{equation}
For zero-velocity initial conditions, this expression yields $\dot{Q}(0) =0$, in contrast to
the time series depicted in figure~\ref{3tre}b. 
This discrepancy brings forth a limitation of the hydrostatic (and hence dispersionless) long-wave model. 
It is generally accepted that the dispersionless approximation works well at intermediate times,
while at long times the system could display a gradient catastrophe, which 
can be avoided  by restoring dispersive effects~(Esler and Pearce, 2011). Remarkably,
equation~\req{qtsw} shows that dispersive effects can also be qualitatively relevant at short times, even in the absence of 
large $x$-derivatives. 
 Specifically, at $t=0$ with zero initial velocities the dispersive terms turn~(\ref{qtsw}) into
\begin{equation}
\int_{-L/2}^{+L/2} \frac{-\dot{Q}(0) + \frac{1}{3} \left( \eta_1^3 {\ouu}_{xt}+\eta_2^3 {\oul}_{xt} \right)_x }{\eta_1/\rho_1 +\eta_2/\rho_2}\ dx= 
0.
\label{nonhydro}
\end{equation}
By computing the leading-order long-wave asymptotic expressions for the time derivatives
(Choi \& Camassa, 1999)
in equation~(\ref{nonhydro}), the initial slope of the flux turns out to be
\begin{equation*}
\label{QpAB}
\dot{Q}(0) = \left(\int_{-L/2}^{+L/2} \frac{B_x}{\eta_1/\rho_1 +\eta_2/\rho_2}\ dx\right) 
\left(\int_{-L/2}^{+L/2} \frac{1-A_x}{\eta_1/\rho_1 +\eta_2/\rho_2}\ dx\right)^{-1},\, \text{where}
\end{equation*}
\begin{equation*}
A= 
\frac{\eta_1^3}{3} \Big( \frac{\rho_2}{\eta_2\rho_1 + \eta_1\rho_2}  \Big)_x +(1\leftrightarrow 2),
\quad
B=
\frac{g (\rho_2-\rho_1) \eta_1^3}{3} \Big(\frac{\eta_2 {\eta_2}_x }{\eta_2\rho_1 + \eta_1\rho_2} \Big)_x 
-(1\leftrightarrow 2)\, .
\end{equation*}
Even within this leading-order approximation, there is rough agreement (in particular by capturing the correct sign) with the numerical data in figure~\ref{3tre}b. This can also be seen as an a posteriori check on the robustness of the two-layer model. For instance,  the theoretical prediction (adjusting for 
smooth stratification, as in Camassa \& Tiron, 2011) is $\dot{Q}(0) \simeq -8.1\times 10^{-3} $ cm$^{2}$/s$^{2}$ for 
the case in figure~\ref{3tre}b with $L=1232$ cm, whereas the numerical result is $\dot{Q}(0) \simeq -1.9 \times 10^{-3}$ cm$^{2}$/s$^{2}$.
Finally, we remark that 
the inertia effects can be further magnified by taking larger density variations. We have carried out tests
with various density ratios, e.g., for $\rho_2=2\rho_1$ 
and $\rho_2=1.022$ g/cm$^3$ the model predicts $\dot{Q}(0) \simeq -9.62$ cm$^2$/s$^{2}$, while the measured numerical value is $\dot{Q}(0) \simeq -2.04$~cm$^2$/s$^{2}$. 

\section*{Acknowledgments}
Partial support by NSF grants DMS-0509423, DMS-1009750, RTG DMS-0943851 and CMG ARC-1025523, as well as by the 
MIUR Cofin2008 project  20082K9KXZ 
is acknowledged.
{R.C., S.C. and M.P. thank  the {\em Dipartimento di Matematica e Applicazioni\/} of the 
Milano-Bicocca University for hospitality.} 
We thank P. Milewski for sending us, while this work was being completed, the preprint of his paper with A. Boonkasame, and  the referees for 
providing valuable feedback on the manuscript's first version. 

\singlespacing

\end{document}